\documentstyle[twocolumn,prl,tighten,aps,amstex,graphicx,times,floats]{revtex}

\begin{document}

\thispagestyle{empty}

% macros for marking changes
\marginparwidth 1.cm
\setlength{\hoffset}{-1cm}
\newcommand{\mpar}[1]{{\marginpar{\hbadness10000%
                      \sloppy\hfuzz10pt\boldmath\bf\footnotesize#1}}%
                      \typeout{marginpar: #1}\ignorespaces}
\def\mda{\mpar{\hfil$\downarrow$\hfil}\ignorespaces}
\def\mua{\mpar{\hfil$\uparrow$\hfil}\ignorespaces}
\def\mla{\marginpar[\boldmath\hfil$\rightarrow$\hfil]%
                   {\boldmath\hfil$\leftarrow $\hfil}%
                    \typeout{marginpar: $\leftrightarrow$}\ignorespaces}

\renewcommand{\abstractname}{Abstract}
\renewcommand{\figurename}{Figure}
\renewcommand{\refname}{Bibliography}

% peter's conventions 
\newcommand{\eg}{{\it e.g.}\;}
\newcommand{\ie}{{\it i.e.}\;}
\newcommand{\etal}{{\it et al.}\;}
\newcommand{\ibid}{{\it ibid.}\;}

% additional commands 
\newcommand{\mx}{M_{\rm SUSY}}
\newcommand{\pt}{p_{\rm T}}
\newcommand{\et}{E_{\rm T}}
\newcommand{\del}{\varepsilon}
\newcommand{\sla}[1]{/\!\!\!#1}

% some journals 
\newcommand{\zpc}[3]{${\rm Z. Phys.}$ {\bf C#1} (#2) #3}
\newcommand{\epc}[3]{${\rm Eur. Phys. J.}$ {\bf C#1} (#2) #3}
\newcommand{\npb}[3]{${\rm Nucl. Phys.}$ {\bf B#1} (#2)~#3}
\newcommand{\plb}[3]{${\rm Phys. Lett.}$ {\bf B#1} (#2) #3}
\renewcommand{\prd}[3]{${\rm Phys. Rev.}$ {\bf D#1} (#2) #3}
\renewcommand{\prl}[3]{${\rm Phys. Rev. Lett.}$ {\bf #1} (#2) #3}
\newcommand{\prep}[3]{${\rm Phys. Rep.}$ {\bf #1} (#2) #3}
\newcommand{\fp}[3]{${\rm Fortschr. Phys.}$ {\bf #1} (#2) #3}
\newcommand{\nc}[3]{${\rm Nuovo Cimento}$ {\bf #1} (#2) #3}
\newcommand{\ijmp}[3]{${\rm Int. J. Mod. Phys.}$ {\bf #1} (#2) #3}
\renewcommand{\jcp}[3]{${\rm J. Comp. Phys.}$ {\bf #1} (#2) #3}
\newcommand{\ptp}[3]{${\rm Prog. Theo. Phys.}$ {\bf #1} (#2) #3}
\newcommand{\sjnp}[3]{${\rm Sov. J. Nucl. Phys.}$ {\bf #1} (#2) #3}
\newcommand{\cpc}[3]{${\rm Comp. Phys. Commun.}$ {\bf #1} (#2) #3}
\newcommand{\mpl}[3]{${\rm Mod. Phys. Lett.}$ {\bf #1} (#2) #3}
\newcommand{\cmp}[3]{${\rm Commun. Math. Phys.}$ {\bf #1} (#2) #3}
\newcommand{\jmp}[3]{${\rm J. Math. Phys.}$ {\bf #1} (#2) #3}
\newcommand{\nim}[3]{${\rm Nucl. Instr. Meth.}$ {\bf #1} (#2) #3}
\newcommand{\prev}[3]{${\rm Phys. Rev.}$ {\bf #1} (#2) #3}
\newcommand{\el}[3]{${\rm Europhysics Letters}$ {\bf #1} (#2) #3}
\renewcommand{\ap}[3]{${\rm Ann. of~Phys.}$ {\bf #1} (#2) #3}
\newcommand{\jhep}[3]{${\rm JHEP}$ {\bf #1} (#2) #3}
\newcommand{\jetp}[3]{${\rm JETP}$ {\bf #1} (#2) #3}
\newcommand{\jetpl}[3]{${\rm JETP Lett.}$ {\bf #1} (#2) #3}
\newcommand{\acpp}[3]{${\rm Acta Physica Polonica}$ {\bf #1} (#2) #3}
\newcommand{\science}[3]{${\rm Science}$ {\bf #1} (#2) #3}
\newcommand{\vj}[4]{${\rm #1~}$ {\bf #2} (#3) #4}
\newcommand{\ej}[3]{${\bf #1}$ (#2) #3}
\newcommand{\vjs}[2]{${\rm #1~}$ {\bf #2}}
\newcommand{\hep}[1]{${\tt hep\!-\!ph/}$ {#1}}
\newcommand{\hex}[1]{${\tt hep\!-\!ex/}$ {#1}}
\newcommand{\desy}[1]{${\rm DESY-}${#1}}
\newcommand{\cern}[2]{${\rm CERN-TH}${#1}/{#2}}

\preprint{
\font\fortssbx=cmssbx10 scaled \magstep2
\hbox to \hsize{
\hskip.5in \raise.1in\hbox{\fortssbx University of Wisconsin - Madison}
\hfill\vtop{\hbox{\bf MADPH-01-1229}
            \hbox{\bf FERMILAB-Pub-01/067-T}
            \hbox{hep-ph/0105325}
            \hbox{May 2001}} }
}

\title{ 
Determining the Structure of Higgs Couplings at the LHC 
} 

\author{
Tilman Plehn${}^1$, 
David Rainwater${}^{2}$, and 
Dieter Zeppenfeld${}^1$ 
} 

\address{ 
${}^1$
Department of Physics, University of Wisconsin, Madison, WI, USA \\
${}^{2}$
Theory Dept., Fermi National Accelerator Laboratory, Batavia, IL, USA
} 

\maketitle 

\begin{abstract}
Higgs boson production via weak boson fusion at the 
CERN Large Hadron Collider has the capability to determine the dominant 
CP nature of a Higgs boson, via the tensor structure of its 
coupling to weak bosons. This information is contained in the azimuthal 
angle distribution of the two outgoing forward tagging jets. The 
technique is independent of both the Higgs boson mass and the observed 
decay channel.
\end{abstract} 

\vspace{0.2in}

%%%%%%%%%%%%%%%%%%%%%%%%%%%%%%%  MAIN TEXT  %%%%%%%%%%%%%%%%%%%%%%%%%%%%

The CERN Large Hadron Collider (LHC) is generally regarded as a tool 
that can guarantee direct observation of a Higgs boson, the remnant of
the mechanism believed responsible for electroweak symmetry breaking and 
fermion  mass generation, and the last unobserved element of the 
Standard Model (SM) of elementary particle physics. Furthermore, the LHC 
promises complete coverage of Higgs decay scenarios~\cite{tdr+,dreiner}, 
including general MSSM parameterizations~\cite{tdr+,wbf_ll}, and even 
invisible Higgs decays~\cite{wbf_inv}. This capability has been greatly 
enhanced recently by 
the addition of the weak boson fusion (WBF) production channel to the 
search strategies~\cite{wbf_ll,wbf_aa,wbf_ww}.
While being extremely useful at the LHC, WBF has too low a rate and is too 
similar to background processes at the Fermilab Tevatron~\cite{run2higgs}.

Observation of a resonance in some expected decay channel is, however,
only the beginning of Higgs physics. Continuing efforts will include 
the search for more than one Higgs boson, as predicted e.g. by two-Higgs 
doublet models, of which the MSSM~\cite{heavy_susy,higgs_self} is a 
subset. At least as important is the detailed study of the properties 
of the Higgs-like resonance, not only at a future Linear 
Collider~\cite{tesla} but also at the LHC: determination of all the 
quantum numbers and couplings of the state. These include the gauge, 
Yukawa and self-couplings as well as the charge, color, spin, and CP 
quantum numbers. While charge and color identification is 
straightforward and a technique has been proposed for the gauge and 
Yukawa coupling determinations~\cite{couplings}, the LHC has 
considerable difficulty in practice to determine the Higgs CP 
transformation properties for intermediate Higgs masses~\cite{old_cp} 
via a weak boson coupling, and no technique has yet been proposed to 
identify the tensor structure of the Higgs-weak boson vertex in the 
intermediate mass range. The methods of Ref.~\cite{gunion} may be useful, 
but only for very light Higgs masses. Furthermore, this method does not 
involve the weak boson vertices at all. 

In this letter we propose a technique which achieves the CP measurement 
goal via a study of WBF events. WBF Higgs production, while not the largest 
cross section at the LHC, is useful because of its characteristic 
kinematical structure, involving two forward tagging jets and central Higgs 
decay products, which allows one to isolate the signal in a low background 
environment. The angular distribution of the two tagging jets carries 
unambiguous information on the CP properties of the Higgs couplings to 
weak bosons which is independent of the Higgs decay channel observed.

\bigskip

As a theoretical framework we consider two possible ways to couple a
spin zero field to two gauge bosons via higher dimensional 
operators. In a gauge invariant dimension six (D6) Lagrangian, the terms
\begin{equation}
{\cal L}_6 = 
  \frac{g^2}{2\Lambda_{{\rm e}, 6}^2} 
    \left( \Phi^\dagger \Phi \right) 
    V_{\mu\nu} {V}^{\mu\nu}         
+ \frac{g^2}{2\Lambda_{{\rm o}, 6}^2} 
 \left( \Phi^\dagger \Phi \right) 
   \widetilde{V}_{\mu\nu} {V}^{\mu\nu}
\label{eq:d6}
\end{equation}
lead to anomalous couplings between the Higgs-type scalar and two
charged gauge bosons~\cite{lagrangian}. The scales $\Lambda_{\rm e}$
and $\Lambda_{\rm o}$ set the coupling strength of CP even and CP odd 
scalars, respectively.  
The Feynman rules can be read off the dimension five (D5) operators that 
result when $\Phi$ is given a physical field expansion:
\begin{equation}
{\cal L}_5 = 
  \frac{1}{\Lambda_{{\rm e}, 5}} \; H \;
    W^+_{\mu\nu} {W^-}^{\mu\nu}         
+ \frac{1}{\Lambda_{{\rm o}, 5}} \; H \; 
   \widetilde{W}^+_{\mu\nu} {W^-}^{\mu\nu}
\label{eq:d5}
\end{equation}
and similarly for the $Z$ boson.
The two scales are related via $1/\Lambda_5 = g^2 v/\Lambda_6^2$. 
Since we assume SU(2) invariance and do not consider additional D6 operators
like $\Phi^\dagger \Phi \;B_{\mu\nu}B^{\mu\nu}$, the $WWH$ and $ZZH$ 
couplings are related by the same $\cos^2\theta_W$ factor as in the SM. 

In principle one would have to introduce a form factor to ensure the 
unitarity of scattering amplitudes involving these operators. However,
we have checked that at the LHC the typical $p_T$ of the tagging jets,
for WBF processes generated by the D5 operators, remains comparable to 
the SM case and is well below the scale $\Lambda$, which we assume to 
be of order a few hundred GeV or above. Thus, form-factor effects would 
remain small in a more complete treatment and they would not distort 
the angular distributions to be discussed below.

\medskip

The analog of the CP even D5 operators is present in Higgs production 
through gluon fusion, as $HG_{\mu\nu}G^{\mu\nu}$, and gives an excellent 
approximation for the $ggH$ coupling induced by heavy quark (and squark) 
loops. In the low energy limit the D5 operators also appear in the 
one-loop $WWH$ coupling, but their size is suppressed by a factor 
$\alpha_W/\pi \sim 10^{-2}$ and hence not observable at the LHC, as we 
will see later. Another source would be a Higgs-like top-pion that is 
a general feature of topcolor models~\cite{topcolor} and which couples 
to weak bosons like $\Pi\widetilde{W}^+_{\mu\nu} {W^-}^{\mu\nu}$ with a 
coefficient that is considerably larger than in the SM and is expected 
to lead to observable rates of production in the WBF channel.

%One particularly nice feature appears once we require U(1) and SU(2)
%gauge invariance of the higher dimensional operators: the set of D6
%operators in Eq.~(\ref{eq:d6}) would give rise to a direct 
%$\gamma\gamma H$ coupling. This might dramatically enhance the Higgs 
%branching fraction to photons, thus speeding up observation in the WBF
%production mode~\cite{wbf_aa}. In addition, the azimuthal asymmetry, to be 
%defined below, would make it possible to identify 
%the nature of the additional coupling unambiguously~\cite{oscar}.

For a true Higgs boson the $WWH$ and $ZZH$ couplings originate from the 
kinetic energy term of the symmetry breaking field, 
$(D_\mu\Phi)^\dagger (D^\mu\Phi)$, which mediates couplings proportional 
to the metric tensor. This tensor structure is not 
gauge-invariant by itself and identifies the Higgs field as the remnant 
of spontaneous symmetry breaking. It is thus crucial to distinguish it 
from the effective couplings derived from Eq.~(\ref{eq:d6}). 
Since the partons in the WBF processes
\begin{equation}
p p \to q q' H \to q q' \tau \tau, \; q q' WW, \; q q' \gamma \gamma  
\end{equation}
are approximately massless, the production cross section is 
proportional to the Higgs-weak boson coupling squared. Replacing the 
$g^{\mu\nu}$ coupling with a higher dimensional coupling changes the 
kinematical structure of the final state scattered quarks. 

\bigskip

To illustrate this we consider leptonic final states in 
$H\to\tau\tau$ decays as in 
Ref.~\cite{wbf_ll}. We emphasize the $H\to\tau\tau$ decay channel 
because it is resilient to modifications of the Higgs sector as 
encountered in the MSSM: a luminosity of 40~fb${}^{-1}$ 
guarantees coverage of the entire $(m_A$-$\tan \beta)$ plane after 
combining the leptonic and semileptonic decay channels of the tau 
pair~\cite{wbf_ll}. 
The basic set of cuts on the outgoing partons consists of 
\begin{eqnarray}
p_{T_j} \geq 20~{\rm GeV} \qquad 
\triangle R_{jj} \geq 0.6 \qquad
|\eta_j| \leq 4.5 \notag \\
|\eta_{j_1}-\eta_{j_2}| \geq 4.2 \qquad
\eta_{j_1} \cdot \eta_{j_2} < 0
\label{eq:cuts}
\end{eqnarray}
in addition to the separation and acceptance cuts for the decay leptons, 
which we don't discuss here. (Further cuts on the invariant mass of the 
tagging jets and the tau pair decay kinematics are necessary to extract 
the signal. These details and the final step of reconstructing the tau 
pair invariant mass are currently under study by various CMS and ATLAS 
groups, with very encouraging results~\cite{wbf_exp}.)
In the parton level analysis we are left with a cross section of 
$\sigma \sim 0.5$~fb for a 120~GeV SM Higgs boson, leading to $S/B=2.7/1$
and a Gaussian significance $\sigma_{\rm Gauss} = 6.8$ for 60~fb$^{-1}$ of
data~\cite{wbf_ll}. 
The two largest backgrounds are QCD and electroweak $\tau\tau jj$ 
production, which together are $\lesssim 30\%$ of the signal cross 
section after cuts. The other backgrounds, including $H \to WW$ and 
$t\bar{t}+$~jets, are of minor importance and can safely be neglected 
in the following qualitative analysis.

\bigskip \bigskip

\begin{figure}[t] 
\begin{center}
\includegraphics[width=9.0cm]{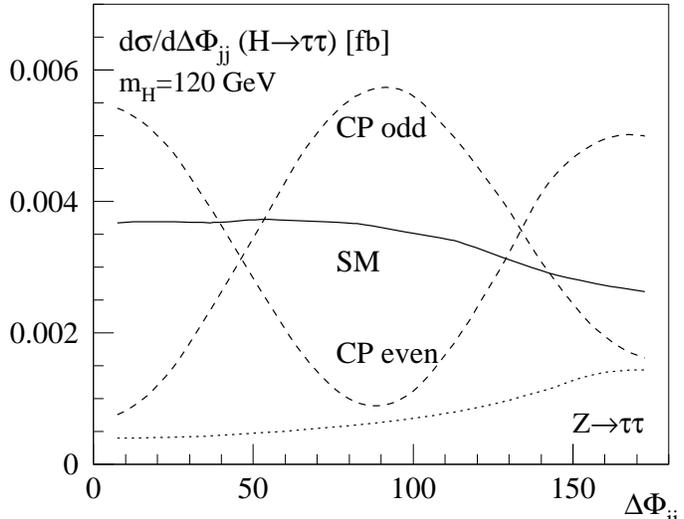}
\caption[]{\label{fig:squared} 
  Azimuthal angle distribution between the two tagging jets  
  for the signal and dominant $\tau\tau$ backgrounds, $m_H = 120$~GeV. 
  Cross sections for the D5 operators correspond to $\Lambda_5=480$~GeV,
  which reproduces the SM cross section,
  after cuts as in Eq.(\ref{eq:cuts}) and 
  Ref.~\protect\cite{wbf_ll}. 
  The expected SM background is added to all three Higgs curves.}
\end{center}
\end{figure}

\begin{figure}[t] 
\begin{center}
\includegraphics[width=9.0cm]{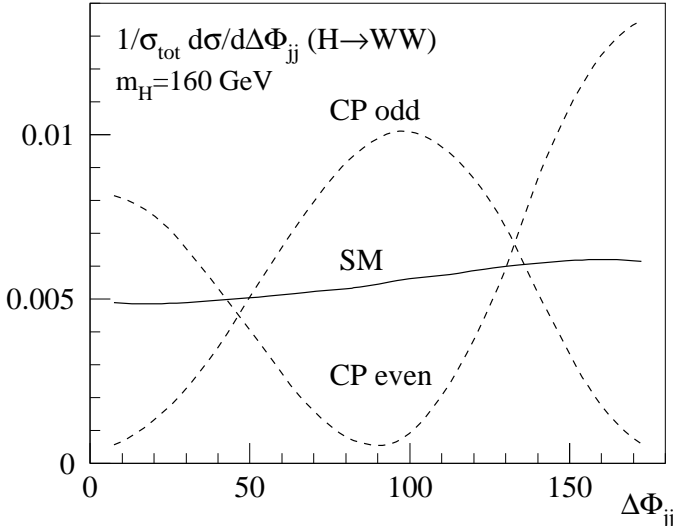}
\caption[]{\label{fig:squared_ww} 
  Normalized distributions of the azimuthal angle between the two 
  tagging jets, for the $H\to WW\to e\mu\sla p_T$ signal at $m_H = 160$~GeV. 
  Curves are for the SM and for single D5 operators as given in 
  Eq.~(\ref{eq:d5}), after cuts as in Eq.(\ref{eq:cuts}) and 
  Ref.~\protect\cite{wbf_ww}. 
}
\end{center}
\end{figure}

{\bf 1.}
Let us first assume that a Higgs-like scalar signal is found at the LHC 
in this channel at the expected SM rate. We must experimentally 
distinguish a SM $g^{\mu\nu}$-type coupling from the tensor structures 
implied by the D5 operators of Eq.~(\ref{eq:d5}). A SM rate induced by 
one of the D5 operators requires a scale 
$\Lambda_5 \approx 480$~GeV ($\Lambda_6 \approx 220$~GeV). 
A particularly interesting kinematic variable is the azimuthal angle 
$\Delta \phi_{\rm jj}$ between the two tagging jets. For forward 
scattering, which is dominant due to the $W$-propagator factors, the 
remaining SM matrix element squared for $qq\to qqH$ is proportional to 
$\hat s\; m_{jj}^2$, where $m_{jj}$ is the invariant mass of the two 
tagging jets. This leads to an essentially flat azimuthal angle 
distribution between the two jets, as shown in Figs.~\ref{fig:squared}
and ~\ref{fig:squared_ww}. In the $H\to\tau\tau$ case, a slight bias 
toward small angles is introduced by selection cuts, 
which require a substantial transverse momentum for the Higgs boson. 
The major backgrounds, $Zjj$ production with $Z\to\tau\tau$, possess 
mostly back-to-back tagging jets.

For the CP odd D5 operator, the shape of the distribution follows from 
the presence of the Levi-Civita tensor 
in the coupling: it gives a nonzero result only if there are four 
independent momenta in the process (here, the four external 
parton momenta). For planar events, i.e. for tagging jets which are 
back-to-back or collinear in the transverse plane, the matrix element 
vanishes.

The CP even operator given in Eq.(\ref{eq:d5}) develops a special 
feature for forward tagging jets. In the limit of 
$|p^{\rm (tag)}_z| \gg |p^{\rm (tag)}_{x,y}|$ and small energy loss
of the two scattered quarks, we can approximate the matrix element by 
\begin{eqnarray}
{\cal M}_{{\rm e}, 5} &\propto& \frac{1}{\Lambda_{{\rm e}, 5}}
                      J_1^\mu J_2^\nu \;
                      \left[ g_{\mu \nu} (q_1 \cdot q_2)
                           - q_{1 \nu} q_{2 \mu} \right] \notag \\
                      &\sim& \frac{1}{\Lambda_{{\rm e}, 5}}
                      \left[ J_1^0 J_2^0 - J_1^3 J_2^3 \right]
                       {\bf p}^{\rm (tag 1)}_{\rm T} \cdot 
                             {\bf p}^{\rm (tag 2)}_{\rm T}   
\label{eq:matrix}
\end{eqnarray}
where $q_i,J_i$ are the momenta and currents of the intermediate weak 
gauge bosons. For $\Delta\phi_{\rm jj} = \pi/2$ the last term vanishes, 
leading to an approximate zero in the distribution. From the three 
curves shown in Fig.~\ref{fig:squared} we conclude that the azimuthal 
angle distribution is a gold plated observable for determining the 
dominant CP nature and the tensor structure of the Higgs coupling. 
With 100~fb$^{-1}$ of data per experiment, the SM case 
can be distinguished from the CP even (CP odd) D5 couplings with a 
statistical power of $\sim 5$ ($4.5$) sigma, from the 
$H\to\tau\tau$ channels~\cite{wbf_ll} alone. 
{\it This observable is furthermore independent of the particular 
decay channel and Higgs mass range.} 
We have explicitly checked the case of a 160~GeV Higgs boson decaying 
to $W$ pairs and find exactly the same features, shown in 
Fig.~\ref{fig:squared_ww}. Note, however, that in this case decay 
distributions will depend on the structure of the $HWW$ vertex also. 
%No reconstruction or 
%manipulation is needed beyond what is required to extract the WBF signal 
%in the first place.

\bigskip \bigskip

{\bf 2.}
Let us now examine the following scenario: a Higgs candidate is found 
at the LHC with a predominantly Standard Model $g^{\mu\nu}$ coupling. 
How sensitive will experiments be to any additional D5 contribution?

\medskip

For the CP odd D5 coupling we do not observe any interference term 
between the Standard Model and the D5 matrix element. Although there is 
a non-zero contribution at the matrix element level, any hadron collider 
observable is averaged over charge conjugate processes since we cannot 
distinguish quark from antiquark jets. As a result, interference effects
cancel in typical hadronic differential cross sections. 
Using the azimuthal angle distribution will only marginally enhance 
the sensitivity to a small contribution of the CP odd Higgs coupling 
beyond what a measurement of the Higgs production cross section could
give.

\medskip

\begin{figure}[t] 
\begin{center}
\includegraphics[width=9.0cm]{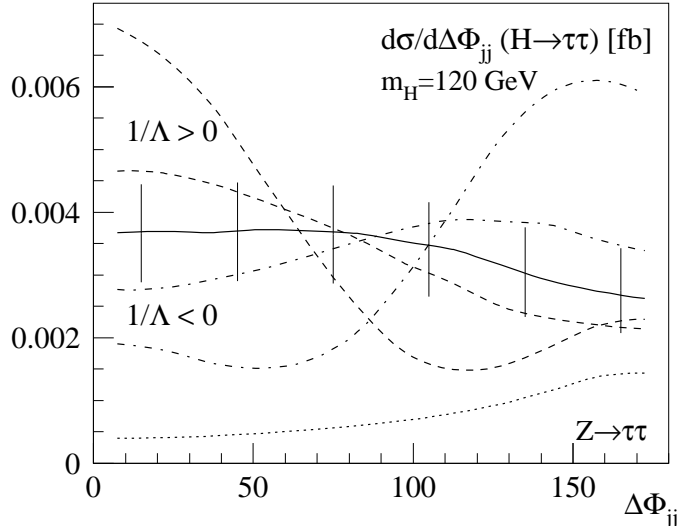}
\caption[]{\label{fig:inter} 
  Azimuthal jet angle distribution for the SM  and 
  interference with a CP even D5 coupling. The two 
  curves for each sign of the operator correspond to values 
  $\sigma/\sigma_{\rm SM}=0.04,1.0$. Error bars 
  for the signal and the dominant backgrounds correspond to an integrated 
  luminosity of 100~fb${}^{-1}$ per experiment, distributed over 6 bins,
  and are statistical only.}
\end{center}
\end{figure}

In the case of a contribution from a CP-even D5 operator, interference 
effects are important for the distortion of the $\phi_{\rm jj}$ 
distribution. All additional terms in the squared amplitude 
$|{\cal M}|^2 = |{\cal M}_{\rm SM} + {\cal  M}_{{\rm e}, 5}|^2$ have an 
approximate zero at $\Delta \phi_{\rm jj} = \pi/2$, according to 
Eq.(\ref{eq:matrix}). Moreover, the dominant piece of the anomalous 
amplitude changes sign at this approximate zero which results in a sign 
change of the interference term at $\pi/2$. Fig.~\ref{fig:inter} shows 
that, dependent on the sign of the D5 operator, the maximum of the 
distribution is shifted to large or small angles $\Delta \phi_{\rm jj}$. 
Results are shown for two different values of the scale $\Lambda_5$ 
which are chosen such that the D5 operator alone, without a SM 
contribution, would produce a Higgs production cross section, $\sigma$, 
which equals 0.04 (1.0) of the SM cross section, $\sigma_{\rm SM}$. 
While changes in cross sections of a few percent are most likely 
beyond the reach of any LHC counting experiment, we see that in the 
differential cross section the effect of D5 operators is quite 
significant~\cite{oscar}.

\begin{figure}[t] 
\begin{center}
\includegraphics[width=9.0cm]{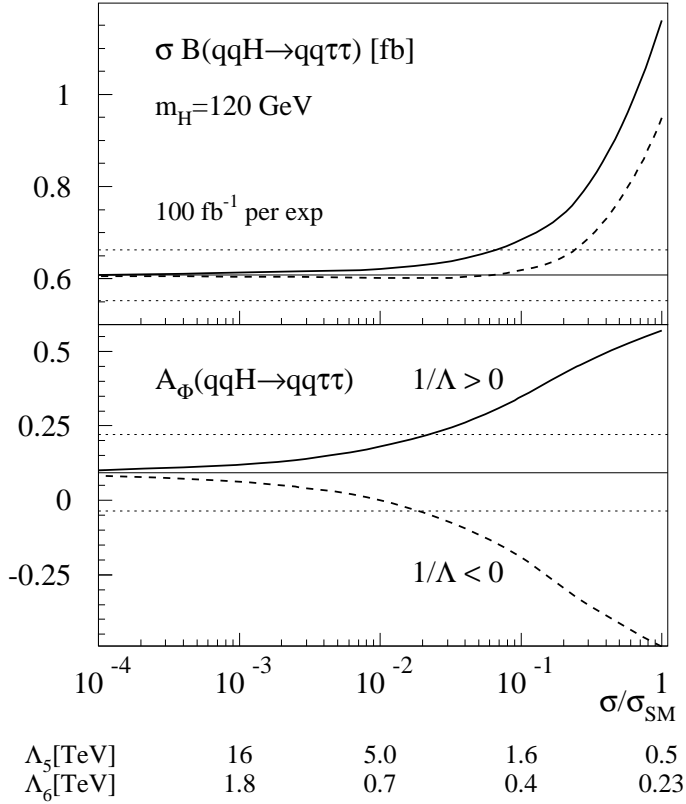}
\vspace*{2mm}
\caption[]{\label{fig:asymmetry} 
  Comparison of the sensitivity of a total cross section (counting) 
  experiment and the azimuthal angle asymmetry, Eq.(\ref{eq:asy}),
  to the presence of a CP even D5 coupling. The 
  horizontal lines represent one sigma statistical deviations from 
  the SM value. The secondary axes show the corresponding values of 
  $\Lambda_5$ and $\Lambda_6$, as defined in Eqs.(\ref{eq:d6},\ref{eq:d5}).}
\end{center}
\end{figure}

\medskip 

To quantify this effect and at the same time minimize systematic
errors we define the asymmetry
\begin{equation}
A_\phi = \frac{ \sigma(\Delta \phi_{\rm jj}<\pi/2)
               -\sigma(\Delta \phi_{\rm jj}>\pi/2)}
              { \sigma(\Delta \phi_{\rm jj}<\pi/2)
               +\sigma(\Delta \phi_{\rm jj}>\pi/2)}\;.
\label{eq:asy}
\end{equation}
One major source of systematic uncertainty will be the gluon fusion 
induced $H+ 2$~jets background, which in the large top mass limit is 
proportional to the CP even D5 operator $HG_{\mu\nu}G^{\mu\nu}$. 
At the amplitude level, this operator induces the same azimuthal angle 
dependence of the two 
jets as the CP even operator of Eq.~(\ref{eq:d5}). However, since it 
contributes to $H+2$~jets via $t$-channel gluon (color octet) exchange, 
it cannot 
interfere with WBF. This gluon fusion contribution can exceed 
${\cal O}(10\%)$ of the signal after cuts~\cite{dieter_carlo} and is 
expected to have large higher order QCD corrections~\cite{loops}. The 
measurement of the absolute rate of WBF events would therefore be 
systematics limited, due to the unknown $K$-factor for the gluon fusion 
contamination. Assuming that this $K$-factor does not vary with 
$\Delta\phi_{jj}$, a full shape analysis of the azimuthal angle distribution
allows to distinguish this noninterfering gluon fusion background from an
interfering D5 $HWW$ coupling: the asymmetry is dominated by the interference
terms. As mentioned before there is a loop induced $WWH$ coupling, but 
it is expected to contribute with size 
$\sigma \sim (\alpha/\pi)^2\;\sigma_{SM}$, beyond the reach of even a linear
collider precision experiment~\cite{tesla}.

\medskip

In Fig.~\ref{fig:asymmetry} we compare the sensitivity to D5 couplings
expected from  the total cross section and the azimuthal asymmetry,
respectively. In the integrated cross section, interference effects between 
the SM $g^{\mu\nu}$ coupling and the CP even D5 coupling largely cancel. 
With 100~fb$^{-1}$ per experiment, a total cross section measurement at the 
LHC is sensitive (at the 1-$\sigma$ level, and considering statistical 
errors only) to $\Lambda_{{\rm e},6}< 510$~GeV ($1/\Lambda>0$) or 290~GeV 
($1/\Lambda<0$). In contrast, $A_\phi$ is a much more sensitive observable, 
and equally sensitive to positive and negative $\Lambda$. For both signs 
of the D6 coupling the reach in the leptonic $\tau\tau$ channel is 
$\sim 690$~GeV, significantly better than the counting experiment. A rough 
estimate shows that, for a 120~GeV Higgs boson, the LHC will be sensitive to 
$\Lambda_6 \sim 1$~TeV, after adding the statistics of both 
$\tau\tau$~\cite{wbf_ll}, the $WW$~\cite{wbf_ww}, and the 
$\gamma\gamma$~\cite{wbf_aa} WBF channels. While this reach is not quite 
competitive with the linear collider analysis~\cite{tesla}, it is by no 
means certain that a linear collider will be built, and our work adds a 
dimension to the LHC Higgs analysis previously thought not possible. 

\bigskip

In summary, the weak boson fusion production process is not only a 
competitive discovery channel for an intermediate mass Higgs boson, it 
also offers the opportunity to unveil the structure of the Higgs field's 
coupling to gauge bosons. Using information obtained with generic weak 
boson fusion cuts for the intermediate-mass Higgs search, one can 
unambiguously determine the CP nature of a Higgs-like scalar: the 
azimuthal angle distribution between the tagging jets clearly 
distinguishes the Standard Model $g^{\mu\nu}$ coupling from a typical 
loop induced CP even or CP odd coupling.
In a search for dimension five operators which interfere 
with the SM $HWW$ coupling, an asymmetry analysis of this azimuthal angle 
distribution improves the reach far beyond what is possible in a 
counting experiment, including the determination of the sign of the 
additional coupling.

%%%%%%%%%%%%%%%%%%%%  ACKNOWLEDGMENTS  %%%%%%%%%%%%%%%%%%%%

\acknowledgements

We want to thank T.~Han and O.~\'Eboli for inspiring discussions. 
This research was supported in part by the University of Wisconsin 
Research Committee with funds granted by the Wisconsin Alumni 
Research Foundation and in part by the U.~S.~Department of Energy 
under Contract No.~DE-FG02-95ER40896.

%%%%%%%%%%%%%%%%%%%%%%%  REFERENCES  %%%%%%%%%%%%%%%%%%%%%%%

\bibliographystyle{plain}

\end{document}